\documentclass[twocolumn,showpacs,prb,superscriptaddress]{revtex4}

\usepackage{graphicx}

\newcommand{\av}{_{\mathrm{av}}}
\newcommand{\nsw}{N_{\mathrm{sweep}}}
\newcommand{\nsa}{N_{\mathrm{samp}}}
\newcommand{\figurewidth}{\columnwidth}

\begin{document}

\title{Nontrivial critical behavior of the free energy in the two-dimensional
Ising spin glass with bimodal interactions}

\author{Helmut G.~Katzgraber}
\affiliation{Theoretische Physik, ETH H\"onggerberg,
CH-8093 Z\"urich, Switzerland}

\author{L.~W.~Lee}
\affiliation{Department of Physics, University of California,
Santa Cruz, California 95064, USA}

\author{I.~A.~Campbell}
\affiliation{Laboratoire des Collo\"ides, Verres et Nanomateriaux, 
Universit\'e Montpellier II, 34095 Montpellier, France}


\date{\today}

\begin{abstract}
A detailed analysis of Monte Carlo data on the two-dimensional Ising spin
glass with bimodal interactions shows that the free energy of the model has a
nontrivial scaling. In particular, we show by studying the correlation length
that much larger system sizes and lower temperatures are required to see the
true critical behavior of the model in the thermodynamic limit. Our results 
agree with data by Lukic {\em et al.}~in that the degenerate ground state 
is separated from the first excited state by an energy gap of $2J$.
\end{abstract}

\maketitle

\section{Introduction}
\label{sec:introduction}

The two-dimensional Edwards-Anderson Ising spin glass with bimodal
interactions\cite{edwards:75} has been widely used to study properties of
spin-glass systems. Despite the fact that the model only orders at zero
temperature, its popularity can be ascribed mainly to its ease of 
implementation and simplicity. There has been a long-standing controversy
regarding the behavior of the free energy of the model, in particular the
size of the excitation gap and the critical behavior in the
thermodynamic limit.

The exponential scaling of the free energy, and thus correspondingly of all
other thermodynamic quantities, was first proposed by Wang and
Swendsen.\cite{wang:88} They surmised that 
\begin{equation}
C_{\rm V} \sim \beta^{2} e^{-A \beta J} \; ,
\end{equation}
where $\beta = 1/T$ represents the inverse temperature, $J$ is the magnitude
of the bonds, and $A$ is a numerical prefactor.
Simple analytical arguments for Ising systems with bimodal interactions on
square lattices with coordination 4 suggest that the energy gap should be 
$4J$, i.e., $A = 4$. In addition, according to hyperscaling, the singular 
part of the free energy scales as $\xi^{-d}$ with $d = 2$ and so, 
if $C_{\rm V}$ scales exponentially, we expect that the correlation length 
$\xi$ scales as
\begin{equation}
\xi \sim e^{n\beta J}
\label{eq:expscale}
\end{equation}
with $A = 2 n$, as predicted first by Saul and Kardar.\cite{saul:93}

Wang and Swendsen were the first to calculate numerically
the specific heat of the model and find $A = 2$, thus showing a nontrivial
scaling of the free energy; however their results for small system sizes and few
disorder realizations implied strong corrections to scaling. 
These results were later supported by numerical work of 
Lukic {\em et al.}\cite{lukic:04}~who computed the specific heat of the model
for intermediate system sizes ($L \le 50$) and also showed that $A = 2$. 

The aforementioned results stand in contrast to work by Saul and 
Kardar,\cite{saul:93} who argue that $A = 4$. In addition, in 
their original work $n = 2$, 
a behavior later confirmed independently by Houdayer\cite{houdayer:01} 
who studied the finite-size scaling of the Binder ratio,\cite{binder:81} 
as well as Katzgraber and Lee\cite{katzgraber:05b} who studied the finite-size
scaling of the finite-size correlation length directly via Monte Carlo 
simulations. 
Note that exponential scaling of the correlation length means that the
critical exponent $\nu = \infty$ for the two-dimensional spin glass with
bimodal interactions.


In this work, by using an alternate analysis of the data presented in
Ref.~\onlinecite{katzgraber:05b} for the finite-size correlation
length\cite{cooper:82,ballesteros:00} we show that $n$ changes continuously 
with system size. In particular, our analysis shows that $n = 1$ cannot be 
ruled out (in contrast to previous conclusions -- 
see Ref.~\onlinecite{katzgraber:05b}) and 
might be the proper estimate for $n$ in the thermodynamic limit.
In addition, we compute the specific heat for moderate
system sizes directly via Monte Carlo simulations and show that 
$A = 2$, in agreement with data by Lukic {\em et al.}\cite{lukic:04}

In Sec.~\ref{sec:model} we introduce the model and observables and in 
Sec.~\ref{sec:results} we present our results, followed by conclusions in
Sec.~\ref{sec:conclusions}.

\section{Model and Observables}
\label{sec:model}

The Hamiltonian of the two-dimensional Ising spin glass is given by
\begin{equation}
{\cal H} = - \sum_{\langle i,j\rangle} J_{ij} S_i S_j .
\label{eq:ham}
\end{equation}
$S_i= \pm 1$ represent Ising spins and the sum is over nearest neighbors on 
a square lattice with periodic boundary conditions. 
The interactions $J_{ij} \in \{\pm 1\}$ are bimodally distributed.
For the Monte Carlo simulations we use a combination of single-spin flips,
parallel tempering updates,\cite{hukushima:96,marinari:98b} and rejection-free 
cluster moves\cite{houdayer:01} to speed up equilibration. 
Equilibration of the method is tested by performing a logarithmic data
binning of all observables, and we require that the last three bins agree within
error bars and are independent of the number of Monte Carlo sweeps $\nsw$.
The parameters of the simulation are listed in Table \ref{simparams}. 
\begin{table}
\caption{
Parameters of the simulations. $\nsa$ represents the number of disorder
realizations computed; $\nsw$ is the total number of Monte Carlo sweeps
of the $2 N_T$ replicas for a single sample. $N_T$ is the number of
temperatures in the parallel tempering method and $T_{\rm min}$ represents the
lowest temperature simulated. 
\label{simparams}
}
\begin{tabular*}{\columnwidth}{@{\extracolsep{\fill}} c r r r l }
\hline
\hline
$L$  &  $\nsa$  & $\nsw$ & $T_{\rm min}$ & $N_{\rm T}$  \\
\hline
 32 & $ 5\; 000 $ & $2.0 \times 10^6$ & 0.050 & 20 \\
 48 & $ 1\; 000 $ & $2.0 \times 10^6$ & 0.050 & 20 \\
 64 & $     500 $ & $4.2 \times 10^6$ & 0.200 & 39 \\
 96 & $     609 $ & $6.5 \times 10^6$ & 0.200 & 63 \\
128 & $     420 $ & $2.0 \times 10^6$ & 0.396 & 50 \\
\hline
\hline
\end{tabular*}
\end{table}

The finite-size correlation
length\cite{cooper:82,kim:94,palassini:99b,ballesteros:00,lee:03,katzgraber:04} 
$\xi_L$ is given by
\begin{equation}
\xi_L = {1 \over 2 \sin (|{\bf k}_\mathrm{min}|/2)}
\left[{\chi_{\mathrm{SG}}(0) \over
\chi_{\mathrm{SG}}({\bf k}_\mathrm{min})} - 1 \right]^{1/2} \; ,
\label{eq:xiL}
\end{equation}
where ${\bf k}_\mathrm{min} = (2\pi/L, 0)$ is the smallest nonzero
wave vector, and $\chi_{\mathrm{SG}}({\bf k})$ is the  wave-vector-dependent
spin-glass susceptibility:
\begin{equation}
\chi_{\mathrm{SG}}({\bf k}) = {1 \over N} \sum_{i,j} [\langle S_i S_j
\rangle^2 ]\av e^{i {\bf k}\cdot({\bf R}_i - {\bf R}_j) } \; .
\label{eq:chik}
\end{equation}
In the previous equation $[\cdots]_{\rm av}$ represents a disorder average and
$\langle \cdots \rangle$ a thermal average.

In order to compare to the data of Lukic {\em et al}.\cite{lukic:04},~we also
compute the specific heat of the system:
\begin{equation}
C_{\rm V} = \frac{1}{T^2}\left\{ [\langle {\mathcal H}^2 \rangle - 
\langle {\mathcal H}\rangle^2]_{\rm av} \right\} .
\end{equation}

\section{Results}
\label{sec:results}

\begin{figure}
\includegraphics[width=\figurewidth]{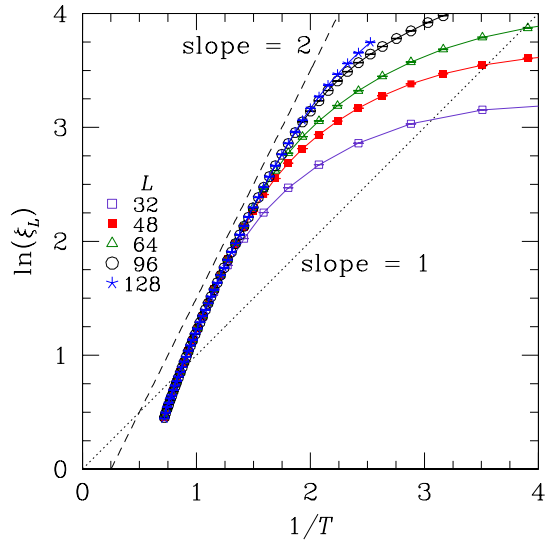}
\vspace*{-1.0cm}
\caption{
(Color online)
Natural logarithm of the finite-size correlation length as a function of $1/T$
for several system sizes (data taken from Ref.~\onlinecite{katzgraber:05b}).
Because $\xi_L \sim \exp(n\beta J)$ we expect the slope of the curves in the
plot to determine $n$. For large $T$ the data are independent of $L$ and
``peel off'' from this common curve at a value of $T$ which decreases with 
increasing size. The slope of the common curve 
asymptotically seems to approach the value $n = 2$ (dashed line). 
The dotted line has slope $n = 1$ which corresponds to the nontrivial scaling.
The data seem incompatible with this prediction.
}
\label{fig:lxil}
\end{figure}

In Fig.~\ref{fig:lxil} we show data for the natural logarithm of 
the finite-size correlation length as a function of $1/T$ for different 
system sizes. We expect data for $\ln(\xi_L)$ vs $1/T$ to approach a slope of
$n$. The data, at first sight, show good agreement with $n = 2$ (dashed line). 
Note that the
data peel off the asymptotic behavior thus masking any potential effective
variation of $n$ with $L$ and $T$. Data at high temperatures show a slope 
$n > 2$, thus suggesting that the slope might change for decreasing
temperatures and increasing system sizes. 

\begin{figure}
\includegraphics[width=\figurewidth]{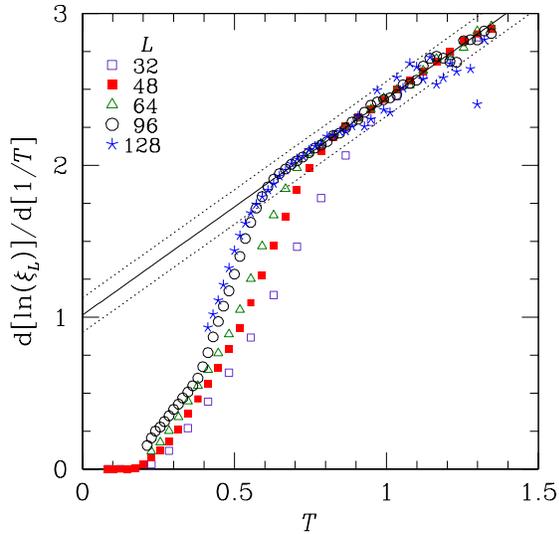}
\vspace*{-1.0cm}
\caption{
(Color online)
Derivative of the natural logarithm of the finite-size correlation length with
respect to the inverse temperature as a function of temperature for several
system sizes. In agreement with Fig.~\ref{fig:lxil} the slope for the
different data sets ranges between 2 and almost 3 for the highest
temperatures, suggesting that the slope $n$ is temperature
dependent. For large temperatures the data are independent of $L$ and well
fitted by a straight line (solid line in plot), thus suggesting that
corrections to $n$ are quadratic. In addition, the data peel off at low
temperatures from the straight-line behavior. An extrapolation of the data to
$T = 0$ agrees well with $n = 1$ in the thermodynamic limit. Note that the
data show that the results in the thermodynamic limit are incompatible 
with $n = 2$ for $L \rightarrow \infty$. The dotted lines correspond to 
error estimates to $n$.
}
\label{fig:slope}
\end{figure}

Figure \ref{fig:slope} shows data for the derivative of the natural logarithm
of the finite-size correlation length with respect to the inverse temperature 
as a function of temperature for several system sizes. In the bulk regime ($T
\gtrsim 0.7$)
where the data show no system size dependence, the computed values for the
slope suggest $2 \lesssim n \lesssim 3$. Interestingly, the data vary linearly
with $T$ and then successively peel off the linear behavior at increasingly
lower temperatures for increasing system sizes. A linear extrapolation of the
data in the bulk regime seems plausible and shows that $n = 1$
at low temperatures in the thermodynamic limit, thus concluding in a nontrivial
scaling of the free energy in the two-dimensional Ising spin glass with
bimodal interactions.

Our results clearly show that the two-dimensional Ising spin glass 
with bimodal interactions has strong
corrections to scaling. In Ref.~\onlinecite{katzgraber:05b} a finite-size
scaling analysis of the finite-size correlation length suggested 
compatibility with $n = 2$, whereas a finite-size scaling of the data for 
$n = 1$ did not seem plausible. Merely the data for the two largest sizes
showed a signature of a scaling behavior when $n = 1$.
This in turn lead to the conclusion that $n = 2$. While the same observable is
being studied here,
the results presented contradict these findings and show that 
to understand the low-temperature properties of this model, 
the limit of $L \rightarrow \infty$ has to be taken {\em before}
extrapolating to $T = 0$. 

\begin{figure}
\includegraphics[width=\figurewidth]{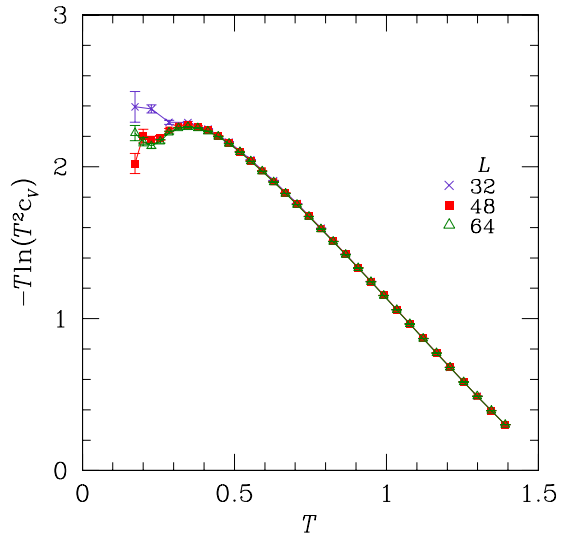}
\vspace*{-1.0cm}
\caption{
(Color online)
Specific heat $C_{\rm V}$ plotted as $-T\ln(T^2C_{\rm V})$ vs $T$ for 
intermediate system sizes. For large enough system sizes $L$ and low 
enough $T$, data for $-T\ln(T^2C_{\rm V})$ should tend to $A$. 
In this case the data agree with the results of Lukic 
{\em et al.}\cite{lukic:04} in that $A \approx 2$ and not $A \approx 4$.
}
\label{fig:cv}
\end{figure}

Finally, in Fig.~\ref{fig:cv} we show data for the specific heat $C_{\rm V}$
as a function of temperature plotted as $-T\ln(T^2C_{\rm V})$ vs $T$. The data
plotted in this way should tend to $A$ for $T \rightarrow 0$. Our results show
that $A \approx 2$ for $T \rightarrow 0$ already for intermediate 
system sizes, in agreement with results by Lukic {\em et al.}\cite{lukic:04}

\section{Conclusions}
\label{sec:conclusions}

To conclude, we have shown data for the correlation length and the specific
heat of the two-dimensional Ising spin glass with bimodal interactions. The
data for the correlation length scale exponentially $\sim \exp(n \beta J)$ and
seem compatible with $n = 1$ in the thermodynamic limit, in contrast to
previous calculations.\cite{katzgraber:05b} A detailed
study of $n$ as a function of temperature shows that $n$ strongly depends on 
system size and temperature. This result is supported by 
data for the specific heat which also scale exponentially with 
$A \approx 2 = 2n$.
This means that the excitation gap for the bimodal spin glass is $\approx 2J$.
Our results in the thermodynamic limit agree with work by 
Lukic {\em et al}.\cite{lukic:04} as well as Wang and Swendsen.\cite{wang:88}

Our data show that corrections to scaling in the two-dimensional Ising spin
glass with bimodal interactions are extremely large. In particular, it is
important to take the limit $L \rightarrow \infty$ before extrapolating to 
$T = 0$. Thus system sizes of at least $\sim 100^2$ and larger are required 
to probe the true thermodynamic behavior. Therefore an analysis with yet 
larger system sizes at lower temperatures is imperative.

\begin{acknowledgments}

We would like to thank F.~Hassler and A.~P.~Young for fruitful 
discussions. 
L.W.L.~acknowledges support from the National Science Foundation under 
NSF Grant No.~DMR 0337049.
The simulations were performed on the Hreidar and Gonzales clusters
at ETH Z\"urich.

\end{acknowledgments}

\bibliography{refs}

\begin{thebibliography}{15}
\expandafter\ifx\csname natexlab\endcsname\relax\def\natexlab#1{#1}\fi
\expandafter\ifx\csname bibnamefont\endcsname\relax
  \def\bibnamefont#1{#1}\fi
\expandafter\ifx\csname bibfnamefont\endcsname\relax
  \def\bibfnamefont#1{#1}\fi
\expandafter\ifx\csname citenamefont\endcsname\relax
  \def\citenamefont#1{#1}\fi
\expandafter\ifx\csname url\endcsname\relax
  \def\url#1{\texttt{#1}}\fi
\expandafter\ifx\csname urlprefix\endcsname\relax\def\urlprefix{URL }\fi
\providecommand{\bibinfo}[2]{#2}
\providecommand{\eprint}[2][]{\url{#2}}

\bibitem[{\citenamefont{Edwards and Anderson}(1975)}]{edwards:75}
\bibinfo{author}{\bibfnamefont{S.~F.} \bibnamefont{Edwards}} \bibnamefont{and}
  \bibinfo{author}{\bibfnamefont{P.~W.} \bibnamefont{Anderson}},
  \emph{\bibinfo{title}{Theory of spin glasses}}, \bibinfo{journal}{J. Phys. F:
  Met. Phys.} \textbf{\bibinfo{volume}{5}}, \bibinfo{pages}{965}
  (\bibinfo{year}{1975}).

\bibitem[{\citenamefont{Wang and Swendsen}(1988)}]{wang:88}
\bibinfo{author}{\bibfnamefont{J.}~\bibnamefont{Wang}} \bibnamefont{and}
  \bibinfo{author}{\bibfnamefont{R.~H.} \bibnamefont{Swendsen}},
  \emph{\bibinfo{title}{{Low-temperature properties of the $\pm J$ spin glass
  in two dimensions}}}, \bibinfo{journal}{Phys. Rev. B}
  \textbf{\bibinfo{volume}{38}}, \bibinfo{pages}{4840} (\bibinfo{year}{1988}).

\bibitem[{\citenamefont{Saul and Kardar}(1993)}]{saul:93}
\bibinfo{author}{\bibfnamefont{L.}~\bibnamefont{Saul}} \bibnamefont{and}
  \bibinfo{author}{\bibfnamefont{M.}~\bibnamefont{Kardar}},
  \emph{\bibinfo{title}{Exact integer algorithm for the two-dimensional
  $\pm{J}$ {I}sing spin glass}}, \bibinfo{journal}{Phys. Rev. E}
  \textbf{\bibinfo{volume}{48}}, \bibinfo{pages}{R3221} (\bibinfo{year}{1993}).

\bibitem[{\citenamefont{Lukic et~al.}(2004)\citenamefont{Lukic, Gallucio,
  Marinari, Martin, and Rinaldi}}]{lukic:04}
\bibinfo{author}{\bibfnamefont{J.}~\bibnamefont{Lukic}},
  \bibinfo{author}{\bibfnamefont{A.}~\bibnamefont{Gallucio}},
  \bibinfo{author}{\bibfnamefont{E.}~\bibnamefont{Marinari}},
  \bibinfo{author}{\bibfnamefont{O.~C.} \bibnamefont{Martin}},
  \bibnamefont{and} \bibinfo{author}{\bibfnamefont{G.}~\bibnamefont{Rinaldi}},
  \emph{\bibinfo{title}{Critical thermodynamics of the two dimensional $\pm j$
  {I}sing spin glass}}, \bibinfo{journal}{Phys. Rev. Lett.}
  \textbf{\bibinfo{volume}{92}}, \bibinfo{pages}{117202}
  (\bibinfo{year}{2004}).

\bibitem[{\citenamefont{Houdayer}(2001)}]{houdayer:01}
\bibinfo{author}{\bibfnamefont{J.}~\bibnamefont{Houdayer}},
  \emph{\bibinfo{title}{A cluster {M}onte {C}arlo algorithm for 2-dimensional
  spin glasses}}, \bibinfo{journal}{Eur. Phys. J. B.}
  \textbf{\bibinfo{volume}{22}}, \bibinfo{pages}{479} (\bibinfo{year}{2001}).

\bibitem[{\citenamefont{Binder}(1981)}]{binder:81}
\bibinfo{author}{\bibfnamefont{K.}~\bibnamefont{Binder}},
  \emph{\bibinfo{title}{Critical properties from {M}onte {C}arlo coarse
  graining and renormalization}}, \bibinfo{journal}{Phys. Rev. Lett.}
  \textbf{\bibinfo{volume}{47}}, \bibinfo{pages}{693} (\bibinfo{year}{1981}).

\bibitem[{\citenamefont{Katzgraber and Lee}(2005)}]{katzgraber:05b}
\bibinfo{author}{\bibfnamefont{H.~G.} \bibnamefont{Katzgraber}}
  \bibnamefont{and} \bibinfo{author}{\bibfnamefont{L.~W.} \bibnamefont{Lee}},
  \emph{\bibinfo{title}{{C}orrelation length of the two-dimensional {I}sing
  spin glass with bimodal interactions}}, \bibinfo{journal}{Phys. Rev. B}
  \textbf{\bibinfo{volume}{71}}, \bibinfo{pages}{134404}
  (\bibinfo{year}{2005}).

\bibitem[{\citenamefont{Cooper et~al.}(1982)\citenamefont{Cooper, Freedman, and
  Preston}}]{cooper:82}
\bibinfo{author}{\bibfnamefont{F.}~\bibnamefont{Cooper}},
  \bibinfo{author}{\bibfnamefont{B.}~\bibnamefont{Freedman}}, \bibnamefont{and}
  \bibinfo{author}{\bibfnamefont{D.}~\bibnamefont{Preston}},
  \emph{\bibinfo{title}{Solving $\phi^4_{1,2}$ theory with {M}onte {C}arlo}},
  \bibinfo{journal}{Nucl. Phys. B} \textbf{\bibinfo{volume}{210}},
  \bibinfo{pages}{210} (\bibinfo{year}{1982}).

\bibitem[{\citenamefont{Ballesteros et~al.}(2000)\citenamefont{Ballesteros,
  Cruz, Fernandez, Martin-Mayor, Pech, Ruiz-Lorenzo, Tarancon, Tellez, Ullod,
  and Ungil}}]{ballesteros:00}
\bibinfo{author}{\bibfnamefont{H.~G.} \bibnamefont{Ballesteros}},
  \bibinfo{author}{\bibfnamefont{A.}~\bibnamefont{Cruz}},
  \bibinfo{author}{\bibfnamefont{L.~A.} \bibnamefont{Fernandez}},
  \bibinfo{author}{\bibfnamefont{V.}~\bibnamefont{Martin-Mayor}},
  \bibinfo{author}{\bibfnamefont{J.}~\bibnamefont{Pech}},
  \bibinfo{author}{\bibfnamefont{J.~J.} \bibnamefont{Ruiz-Lorenzo}},
  \bibinfo{author}{\bibfnamefont{A.}~\bibnamefont{Tarancon}},
  \bibinfo{author}{\bibfnamefont{P.}~\bibnamefont{Tellez}},
  \bibinfo{author}{\bibfnamefont{C.~L.} \bibnamefont{Ullod}}, \bibnamefont{and}
  \bibinfo{author}{\bibfnamefont{C.}~\bibnamefont{Ungil}},
  \emph{\bibinfo{title}{Critical behavior of the three-dimensional {I}sing spin
  glass}}, \bibinfo{journal}{Phys. Rev. B} \textbf{\bibinfo{volume}{62}},
  \bibinfo{pages}{14237} (\bibinfo{year}{2000}).

\bibitem[{\citenamefont{Hukushima and Nemoto}(1996)}]{hukushima:96}
\bibinfo{author}{\bibfnamefont{K.}~\bibnamefont{Hukushima}} \bibnamefont{and}
  \bibinfo{author}{\bibfnamefont{K.}~\bibnamefont{Nemoto}},
  \emph{\bibinfo{title}{Exchange {M}onte {C}arlo method and application to spin
  glass simulations}}, \bibinfo{journal}{J. Phys. Soc. Jpn.}
  \textbf{\bibinfo{volume}{65}}, \bibinfo{pages}{1604} (\bibinfo{year}{1996}).

\bibitem[{\citenamefont{Marinari}(1998)}]{marinari:98b}
\bibinfo{author}{\bibfnamefont{E.}~\bibnamefont{Marinari}},
  \emph{\bibinfo{title}{Optimized {M}onte {C}arlo methods}}, in
  \emph{\bibinfo{booktitle}{Advances in Computer Simulation}}, edited by
  \bibinfo{editor}{\bibfnamefont{J.}~\bibnamefont{Kert\'esz}} \bibnamefont{and}
  \bibinfo{editor}{\bibfnamefont{I.}~\bibnamefont{Kondor}}
  (\bibinfo{publisher}{Springer-Verlag}, \bibinfo{address}{Berlin},
  \bibinfo{year}{1998}), p.~\bibinfo{pages}{50},
  \bibinfo{note}{(cond-mat/9612010)}.

\bibitem[{\citenamefont{{Kim}}(1994)}]{kim:94}
\bibinfo{author}{\bibfnamefont{J.~K.} \bibnamefont{{Kim}}},
  \emph{\bibinfo{title}{{Asymptotic scaling of the mass gap in the
  two-dimensional ${O}(3)$ nonlinear $sigma$ model: {A} numerical study}}},
  \bibinfo{journal}{Phys. Rev. D} \textbf{\bibinfo{volume}{50}},
  \bibinfo{pages}{4663} (\bibinfo{year}{1994}).

\bibitem[{\citenamefont{Palassini and Caracciolo}(1999)}]{palassini:99b}
\bibinfo{author}{\bibfnamefont{M.}~\bibnamefont{Palassini}} \bibnamefont{and}
  \bibinfo{author}{\bibfnamefont{S.}~\bibnamefont{Caracciolo}},
  \emph{\bibinfo{title}{{U}niversal {F}inite-{S}ize {S}caling {F}unctions in
  the 3{D} {I}sing {S}pin {G}lass}}, \bibinfo{journal}{Phys. Rev. Lett.}
  \textbf{\bibinfo{volume}{82}}, \bibinfo{pages}{5128} (\bibinfo{year}{1999}).

\bibitem[{\citenamefont{Lee and Young}(2003)}]{lee:03}
\bibinfo{author}{\bibfnamefont{L.~W.} \bibnamefont{Lee}} \bibnamefont{and}
  \bibinfo{author}{\bibfnamefont{A.~P.} \bibnamefont{Young}},
  \emph{\bibinfo{title}{Single spin- and chiral-glass transition in vector spin
  glasses in three dimensions}}, \bibinfo{journal}{Phys. Rev. Lett.}
  \textbf{\bibinfo{volume}{90}}, \bibinfo{pages}{227203}
  (\bibinfo{year}{2003}).

\bibitem[{\citenamefont{Katzgraber et~al.}(2004)\citenamefont{Katzgraber, Lee,
  and Young}}]{katzgraber:04}
\bibinfo{author}{\bibfnamefont{H.~G.} \bibnamefont{Katzgraber}},
  \bibinfo{author}{\bibfnamefont{L.~W.} \bibnamefont{Lee}}, \bibnamefont{and}
  \bibinfo{author}{\bibfnamefont{A.~P.} \bibnamefont{Young}},
  \emph{\bibinfo{title}{{C}orrelation {L}ength of the {T}wo-{D}imensional
  {I}sing {S}pin {G}lass with {G}aussian {I}nteractions}},
  \bibinfo{journal}{Phys. Rev. B} \textbf{\bibinfo{volume}{70}},
  \bibinfo{pages}{014417} (\bibinfo{year}{2004}).

\end{thebibliography}

\end{document}